\documentclass[12pt,fleqn]{article}

\usepackage{booktabs}
\usepackage{doi}
\usepackage{graphicx}
\usepackage{amssymb,amsmath}
\usepackage{mathrsfs}
\usepackage{color}
\usepackage[margin=1 in]{geometry}

\usepackage{enumitem}
\usepackage{hyperref}

\numberwithin{equation}{section}

\newcommand{\xx}{\tilde{x}}
\newcommand{\yy}{\tilde{y}}
\newcommand{\zz}{\tilde{z}}

\newcommand{\pd}{\partial}

\usepackage{bm}

\begin{document}
	
\title{Integrable systems in magnetic fields: the generalized parabolic cylindrical case}
\author{O. Kub\r{u}$^1$, A. Marchesiello$^2$ and L. \v{S}nobl$^1$}
\maketitle

\begin{center}
{$^1$Czech Technical University in Prague, Faculty of Nuclear Sciences and Physical
	Engineering, Department of Physics, B\v{r}ehov\'{a} 7, 115 19 Prague 1, Czech Republic}\\
{$^2$Czech Technical University in Prague, Faculty of Information Technology,
	Department of Applied Mathematics, Th\'{a}kurova 9, 160 00 Prague 6, Czech Republic} \\[1mm]

E-mail: Ondrej.Kubu@fjfi.cvut.cz, marchant@fit.cvut.cz, Libor.Snobl@fjfi.cvut.cz 
\end{center}

\vspace{10pt}

\begin{abstract}
This article is a contribution to the classification of quadratically integrable systems with vector potentials whose integrals are of the nonstandard, nonseparable type. We focus on generalized parabolic cylindrical case, related to non-subgroup-type coordinates. We find 3 new systems, two with magnetic fields polynomial in Cartesian coordinates and one with unbounded exponential terms. The limit in the parameters of the integrals yields a new parabolic cylindrical system; the limit of vanishing magnetic fields leads to the free motion. This confirms the conjecture that non-subgroup type integrals can be related to separable systems only in a trivial manner.
	\end{abstract}

\vspace{2pc}
\noindent{\it Keywords}: integrability, magnetic field, generalized parabolic cylindrical, nonseparable
\maketitle


	\section{Introduction}
\label{sec:intro}

This article is a step forward to the classification of quadratically integrable and superintegrable systems with vector potentials that possess integrals not connected to separability. We restrict our considerations to the classical nonrelativistic particle on the three-dimensional (3D) Euclidean space $\mathbb{E}_3$ and assume time-independent Hamiltonian.

	Let us recall that an $n$-dimensional classical Hamiltonian system is called \textit{integrable} if it admits $n$ functionally independent first integrals (including the Hamiltonian) that are in involution with respect to the Poisson bracket. If there exist more than $n$ independent integrals, out of which $n$ are in involution, the system is said to be \textit{superintegrable}. \textit{Maximally superintegrable} system has $2n-1$ independent integrals.
	
	In 3D space, integrability implies the existence of two independent integrals $X_1$, $X_2$ in involution in addition to the Hamiltonian, i.e.
\begin{equation}\label{int cond}
	\{X_1,H\}=0, \quad \{X_2,H\}=0, \quad \{X_1,X_2\}=0.
\end{equation}
	
	The solvability of these systems (in quadratures for integrable, algebraically for maximally superintegrable) together with their stability under perturbations makes these systems a bedrock for the study of more complicated and less regular systems. For these reasons, a thorough classification and investigation of their properties merits substantial effort.
	
	Most of the results in the field of superintegrability and the related classification of integrable systems, starting with the seminal papers from the 1960s by Smorodinsky, Winternitz et al.~\cite{Fris1965,Winternitz1967,Makarov1967}, have focused on the so called ``natural'' Hamiltonians on $\mathbb{E}_3$, i.e. Hamiltonians with quadratic but no linear terms in momenta in Euclidean 3D space $\mathbb{E}_3$.
	Classification of such integrable and superintegrable systems with integrals of motion at most quadratic in momenta, based on the 1:1 correspondence between (multi)separable and (super)integrable systems, has been completed on 2D and 3D spaces of constant curvature including $\mathbb{E}_3$, see the review \cite{Miller2013} and the monograph \cite{Miller2018}. Naturally, many articles considering higher order integrals followed, see e.g. \cite{Post2011,Marchesiello2015a,EscobarRuiz2017,EscobarRuiz2018} and references therein. Although systems with only higher order integrals do not separate on the coordinate space, separating coordinates on the phase space for the Post-Winternitz system \cite{Post2011} and its Drach-Holt generalization \cite{CampoamorStursberg2013} were found using the Haantjes algebra approach in \cite{Tempesta2021} and \cite{Reyes2022}, respectively. See also \cite{Kress2023} for the first steps toward the classification of ``natural'' quadratic maximally superintegrable systems in an arbitrary dimension using methods of algebraic geometry.
	
	However, the ``natural'' Hamiltonians certainly do not exhaust all physically relevant systems. For example, the linear terms in the Hamiltonian 
	\begin{equation}\label{HamMagn}
	H=\frac{1}{2}\left(\vec{p}+\vec{A}(\vec{x})\right)^2+W(\vec{x})=\frac{1}{2}\left(\vec{p}\right)^2
	+\vec{p}\cdot \vec{A}(\vec{x})+W(\vec{x})+\frac{1}{2}\left(\vec{A}(\vec{x})\right)^2
\end{equation}
 are necessary for models of a charged particle in an external electromagnetic field or systems in uniform rotation \cite{Zhang}. Despite their relevance, such systems have not been classified yet, probably due to the computational difficulty stemming from the broken 1:1 correspondence between separability and quadratic integrability \cite{Berube2004,McSween2000}. Namely, the separation now requires at least one first-order integral \cite{Shapovalov1972,Benenti2001}. (We note that there is a rather vast literature on magnetic monopole systems in many different spaces, see e.g. \cite{Hoque2018a} for references. However, these are exceptional, particular configurations of the magnetic field, not a classification.)
 
 As a consequence of the broken correspondence, there exist pairs of quadratic integrals on $\mathbb{E}_3$ leading to integrability that do not correspond to any orthogonal separable coordinates \cite{Marchesiello2017}. As the algebraic classification of quadratic terms showed \cite{Marchesiello2022}, there exist classes of integrals with additional quadratic terms extending the original, separable ones, and also with no connection to separability at all. We call such integrals nonstandard. To confirm that these classes are not empty, we have to consider the lower order terms in momenta as well in each class; we have done it for some of them \cite{KMS2022,Hoque2023a,Hoque2023}. Quite a few nonstandard integrable systems were found, though a complete study of all classes has not been achieved yet. The helical undulator in infinite solenoid \cite{KMS2022}, a peculiar system with 3 first order integrals not separable in orthogonal coordinates on the configuration space, remains the sole superintegrable example (also in the relativistic setting \cite{HeinIld}) and is relevant e.g. for the study of free electron lasers \cite{BALAL2020163895}.

In this paper, we consider one of the more complicated classes of \cite{Marchesiello2022}, case (j), which generalizes the parabolic cylindrical structure of integrals by adding leading order terms that depend on three additional parameters.
Our goal is to classify the integrable systems in this class as well as to shed more light on the conjecture in our previous papers \cite{KMS2022,Hoque2023}: In the presence of magnetic fields, there appears to be a discrepancy between the subgroup- and non-subgroup-type coordinates\footnote{The subgroup type coordinates in 3D are defined by the property that some subgroup chain
	\begin{equation}
		E_3 \supset \tilde{G} \supset G_M
	\end{equation}
	exists where $E_3$ is the Euclidean group and $G_M$ is its Abelian subgroup such that the leading order terms of their corresponding quadratic integrals of motion are second--order Casimir operators of subgroups \cite{KMW76,MPW81}.} 
	already at the level of quadratic integrals. 
Namely, all known quadratically integrable systems with vector potentials separate in Cartesian or cylindrical coordinates in the limit of vanishing magnetic field, even when the integrals generalize the leading order structure corresponding to separation in a different coordinate system. In other words, the limit does not lead to separability in non-subgroup coordinates (except the trivial free motion).

The structure of our article is as follows. In Section \ref{sec case j} we introduce our formalism for magnetic fields and the generalized parabolic cylindrical case, i.e. the explicit form of the integrals $X_1$ and $X_2$ depending on parameters $a,b,c$. The following sections contain the detailed classification of integrable systems in this class. Section \ref{sec an0} deals with the case $a\neq0$ and contains one new family, \eqref{B sys aneq0}--\eqref{W sys aneq0}, while Section \ref{sec a0} corresponding to $a=0$ yields 2 more families, \eqref{a0bcn0B}--\eqref{a0bcn0W} and \eqref{B sys a0 exp}--\eqref{W sys a0 exp}. In Section \ref{sec cocl} we summarize our results and consider the limits of vanishing magnetic field as well as vanishing parameters. The latter yields a new standard, not generalized parabolic cylindrical system. In Appendix \ref{app} we present the scalar term in integral $X_1$ for the system \eqref{B sys aneq0}--\eqref{W sys aneq0}, obtained by solving equations \eqref{conds1ord1}.

\section{The generalized parabolic cylindrical case}\label{sec case j}
Before we present the pair of integrals $X_1$ and $X_2$ defining our case, we have to recall a few facts about magnetic fields and introduce our notation.

As we have noted, the Hamiltonian of a system modeling a charged particle in the three-dimensional Euclidean space $\mathbb{E}_3$ immersed in external electromagnetic field is \eqref{HamMagn}, where we have chosen the units $m=1,e=-1$, considering the electron as our main example. We denote the Cartesian coordinates by $\vec{x}=(x,y,z)$ and the corresponding linear momenta $\vec{p}=(p_x,p_y,p_z).$

The key feature of such systems is the invariance of the dynamically relevant magnetic field 
 \begin{equation}
	\vec{B}(\vec{x})=\nabla \times \vec{A}(\vec{x})
\end{equation}
 with respect to the time-independent gauge transformation
\begin{equation}\label{cl gauge}
	\vec A'(\vec{x})= \vec{A} (\vec{x})+\mathrm{d} \chi(\vec{x}), \quad W'(\vec{x})=W(\vec{x}).
\end{equation}

On the other hand, the scalar potential $V(\vec{x})$ (i.e., the momentum-free term in the Hamiltonian~\eqref{HamMagn}) 
\begin{equation}
	V(\vec x)= W(\vec x)+\frac12\left(A_1(\vec x)^2+A_2(\vec x)^2+A_3(\vec x)^2\right)
\end{equation}
is affected by the transformation \eqref{cl gauge}, as well as the first and zero order terms in momenta of quadratic integrals.

In light of that we find it useful to write the integrals of motion in terms of gauge covariant linear and angular momenta
\begin{equation}
	p_i^A=p_i+A_i(\vec{x}),\quad \ell^A_i= \epsilon_{ijk}x_j p^A_k,
\end{equation} 
with $\epsilon_{ijk}$ the totally anti-symmetric Levi-Civita symbol, $\epsilon_{123}=1$. In this way, the conditions coming from \eqref{int cond} become explicitly gauge invariant and depend only on the magnetic field $\vec{B}(\vec{x})$, not on the vector potential $\vec{A}(\vec{x})$.

Let us now introduce case (j) from \cite{Marchesiello2022}: The initial form of the integrals as derived in \cite{Marchesiello2022} is 
\begin{align}
	\nonumber X_1 = {}& \ell^A_x p^A_x+a \ell^A_y p^A_y-(a+1) \ell^A_z p^A_z+\omega \left(\ell^A_x p^A_z -\ell^A_z p^A_x \right) \label{X1or}
	+2 b p^A_x p^A_y+c \left((p^A_y)^2- (p^A_z)^2\right)\\
	{}&+\sum_{j} s_1^{x_j}(\vec{x})p_j^A+m_1(\vec{x}),\\
	X_2 = {}& (p^A_x)^2+ \frac{6 \omega}{4 a-1} p^A_x p^A_z+\frac{a+2}{4 a-1} (p^A_y)^2-\frac{5 a+1}{4 a-1} (p^A_z)^2+\sum_{j} s_2^{x_j}(\vec{x})p_j^A+m_2(\vec{x}),\label{X2or}
\end{align} 
with the parameters
\begin{equation}\label{params}
	\omega = \sqrt{1+a-2 a^2},\quad -\frac{1}{2}< a\leq 0,\quad 0 \leq b,\quad c\in\mathbb{R}.
\end{equation}

Our task is to find the functions $s_i^{x_j}$ and $m_i$ together with the magnetic field $\vec{B}$ and the electrostatic potential $W$ such that these integrals satisfy the integrability conditions \eqref{int cond}.

The integrals contain 3 independent parameters $a,b,c$ and the quantity $\omega$ is nonvanishing within the admissible range of $a$. 
The analysis in \cite{Marchesiello2022} implies that the choice $a=0,\omega=1,c=b=0$ leads to the standard parabolic cylindrical case with integrals $Y_1\equiv \ell_zp_x+ \ldots$ and $Y_2\equiv p_z^2+\ldots$ upon a suitable rotation of the coordinate system and a linear combination of integrals $X_1,X_2$ and $H$. As we are mainly interested in the integrals of the generalized form, we will hereafter assume that at least one of the parameters $a,b,c$ is nonvanishing. As we noted for the elliptic cylindrical case in \cite{Hoque2023}, the determining equations for the standard integrals (i.e. $a=b=c=0$) are less overdetermined and therefore harder to solve. Their solutions include the partially separated (2+1) dimensional system, but the parabolic 2D case has not been completely solved yet \cite{Pucacco_2005}, giving little hope for the full classification in 3D at this moment as well.

The form of the integrals $X_1$ and $X_2$ above was suitable during the algebraic study of the quadratic terms in \cite{Marchesiello2022}, but for our purposes a rotation around the $y$-axis dependent on the parameter $a$,
\begin{equation}\label{rota}
\begin{aligned}
		X ={}& \frac{\omega x}{\psi}+\frac{\psi z}{a +2},\quad Y = y,\quad Z = 
	-\frac{\psi x}{a +2}+\frac{\omega z}{\psi},\\
	{p_X} = {}&\frac{\omega {p_x}}{\psi}+\frac{\psi {p_z}}{a +2}, \quad
	 {p_Y} = {p_y},\quad {p_Z} = -\frac{\psi {p_x}}{a +2}+\frac{\omega p_z}{\psi}, 
\end{aligned}
\end{equation}
where we have defined a real nonvanishing quantity (within our range for $a$)
\begin{equation}
	\psi=\sqrt{-a^2 - a + 2}\label{psi},
\end{equation}
 yields a simpler form with a better physical interpretation.
This transformation is consistent for any $a$ in the range \eqref{params}, including $a=0$, and represents a rotation by the angle $\varphi$ satisfying  $\cos(\varphi)=\frac{\omega}{\psi}$.

In the rotated Cartesian coordinates, which we hereafter denote again by $(x,y,z)$ for simplicity, the integral $X_2$ simplifies significantly if we take its linear combination with the Hamiltonian and rescale it. We have
\begin{align}
	X_1 = {}& -3 a \ell_z^A p_z^A -2 \omega \ell_z^A p_x^A +\frac{a -1}{a +2} c (p_x^A)^{2}+ c (p_y^A)^{2}-\frac{2 a +1}{a +2} c (p_z^A)^{2}
	\label{X1}\\
	{}&-\frac{2 \omega c }{a +2}p_x^A p_z^A+ \frac{2 \psi}{\omega} b p_x^A p_y^A -\frac{2 b \psi }{a +2}p_y^A p_z^A 
	+\sum_{j} s_1^{x_j}(\vec{x})p_j^A+m_1(\vec{x}),\nonumber \\
	X_2 = {}&(p_z^A)^2+\sum_{j} s_2^{x_j}(\vec{x})p_j^A+m_2(\vec{x})\label{X2},
\end{align} 
where the lower order terms were without any loss of generality redefined appropriately after the rotation.

Given the form of the first term in $X_1$, it is convenient to treat the cases with $a\neq0$ and $a=0$ separately, as we shall do in the following two sections.

\section{Parameter $a$ nonvanishing}\label{sec an0}
When $a\neq0$, we simplify the integrals by an additional shift of coordinates,
\begin{equation}\label{shift}
	\xx =x+ \frac{2 b \psi}{3 a \left(a +2\right)},\quad \yy =y- 
	\frac{2 c \omega}{3 a \left(a +2\right)},\quad \zz = z,
\end{equation}
which eliminates some of the mixed terms in $X_1$ while leaving $X_2$ intact (up to redefinition of lower order terms)
\begin{align}
	X_1 = {}& -3 a \ell_{\zz}^A p_{\zz}^A -2 \omega l_{\zz}^A p_{\xx}^A-\frac{\left(5 a^{2}-a -4\right) c (p_{\xx}^A)^{2}}{3 a(a+2)}+ c (p_{\yy}^A)^{2}\nonumber\\
	{}&- \frac{2 a +1}{a+2} c (p_{\zz}^A)^{2}-\frac{2 b \omega\psi}{3a(a-1)}p_{\xx}^A p_{\yy}^A
	+s_1^{\xx} p_{\xx}+s_1^{\yy} p_{\yy}+s_1^{\zz} p_{\zz}+m_1,\label{X1 shift}
\end{align} 
where $s_1^{x_i}$ and $m_1$ are now functions of $(\xx,\yy,\zz)$.

We proceed by the standard direct method, i.e. we impose the Poisson commutativity \eqref{int cond},
separate the equations into different powers in momenta $p_i$, e.g. $p_i p_j$, and impose that all the coefficients vanish. In this way we obtain an overdetermined system of PDEs determining the form of the integrals of motion $X_1,X_2$ as well as the magnetic field $\vec{B}$ and electrostatic potential $W$, hence we call them determining equations.

We present the equations corresponding to the quadratic terms in momenta in \eqref{int cond}, first for the integral $X_1$
\begin{subequations}	\label{conds2ord1}
	\begin{align}
	&\pd_{\xx} s_1^{\xx}=3a \yy B^{\yy}+2\omega\left(
	\xx+\frac{b\psi}{3a(a-1)}\right)B^{\zz},\label{conds2ord11}\\
	 &\pd_{\yy}s_1^{\yy}=3a\xx B^{\xx} -2\omega\left(\xx+\frac{b\psi}{3a(a-1)}\right)B^{\zz}, \label{conds2ord12}\\
	&\pd_{\zz} s_1^{\zz}=-3a (\xx B^{\xx}+\yy B^{\yy}),\label{conds2ord13}\\
	&\pd_{\xx}s_1^{\yy}+\pd_{\yy}s_1^{\xx}=-3a(\yy B^{\xx}+\xx B^{\yy})+\left(4 \omega \yy -\frac{2 c \left(8 a^{2}+5 a -4\right)}{3a (a+2)}\right)B^{\zz},\label{conds2ord14}\\
	&\pd_{\xx}s_1^{\zz}+\pd_{\zz}s_1^{\xx}=-2\omega\left(\xx +\frac{ b \psi}{3 a \left(a -1\right)}\right)B^{\xx}-2\left(2 \omega \yy +\frac{c \left(a +2\right)}{3 a}
	\right)B^{\yy}+3 a \xx B^{\zz},\label{conds2ord15}\\
	&\pd_{\yy}s_1^{\zz}+\pd_{\zz}s_1^{\yy}=\frac{6 c (a+1)}{a+2} B^{\xx}+2\omega\left(\xx+\frac{b \psi}{3 a (a-1)}\right)B^{\yy}+3 a\yy B^{\zz}\label{conds2ord16},
\end{align}
\end{subequations}
and for the integral $X_2$
\begin{equation}
	\begin{aligned}
		&\partial_{\xx}s_2^{\xx}=0,\quad \partial_{\yy}s_2^{\yy}=0,\quad \partial_{z}s_2^{\yy}=0, \\
	&\partial_{\yy}s_2^{\xx}+\partial_{\xx}s_2^{\yy}=0,\\ 
	&\partial_{\zz}s_2^{\yy}+ \partial_{\yy}s_2^{\zz}=-2B^{\xx},\\
	&\partial_{\zz}s_2^{\xx}+\partial_{\xx}s_2^{\zz}=2B^{\yy}.
	\end{aligned} \label{conds2ord2}
\end{equation}

The equations corresponding to the involutivity condition $\{X_1,X_2\}=0$ are rather long, we thus introduce them later when they are needed and some of the functions above already get reduced by a partial solution.

Regardless of the shifted coordinates \eqref{shift}, equations \eqref{conds2ord2} have a standard form and have therefore been solved e.g. in \cite{Fournier2019}. We add to these equations also the condition coming from Maxwell equations,
\begin{equation}\label{divB}
	\nabla\cdot \vec{B}=\pd_{\xx}B^{\xx}+\pd_{\yy}B^{\yy}+\pd_{\zz}B^{\zz}=0,
\end{equation}
which we solve for $\pd_z B^{\zz}.$ We arrive at
\begin{equation}\label{BS2}
	\begin{split}
		B^{\xx}={}&\tfrac{1}{2}(\pd_{\zz}S^{\xx}_{21}(\zz)\xx-\pd_{\yy}S^{\zz}_{2}(\xx,\yy)-\pd_{\zz}S^{\yy}_2(\zz)),\\ B^{\yy}={}&\tfrac{1}{2}(\pd_{\zz}S^{\xx}_{21}(\zz)\yy+\pd_{\xx}S^{\zz}_{2}(\xx,\yy)+\pd_{\zz}S^{\xx}_{22}(\zz)),\\
		B^{\zz}={}&-S^{\xx}_{21}(\zz)+B^{\zz}_{1}(\xx,\yy),\\ s_2^{\xx}={}&S^{\xx}_{21}(\zz)\yy+S^{\xx}_{22}(\zz),\quad s_2^{\yy}=-S^{\xx}_{21}(\zz)\xx+S^{\yy}_2(\zz), \quad s_2^{\zz}=S^{\zz}_{2}(\xx,\yy).
			\end{split}
\end{equation}

We next look at the involutivity equation $\{X_1,X_2\}=0$.
The coefficient of $p_{\xx}^2$ gives 

\begin{equation}
-3a^2 (a-1)\yy^2 \pd_{\zz} S^{\xx}_{21}(\zz)-3a^2(a-1)\yy \pd_{\zz}S^{\xx}_{22}(\zz)+2\omega \left(a(a-1)S^{\yy}_2(\zz)+\frac{b\psi}{3}S^{\xx}_{21}(\zz)\right)=0,
\end{equation}
which implies that all three functions $S^{\xx}_{21}(\zz),S^{\xx}_{22}(\zz)$ and $S^{\yy}_{2}(\zz)$ are constants.
Combining this with the coefficients of $p_{\yy}^2$ and $p_{\xx}p_{\yy}$, we arrive at
\begin{equation}
	S^{\xx}_{21}(\zz)=0,\quad S^{\yy}_{2}(\zz)=0,\quad	S^{\xx}_{22}(\zz)=0.
\end{equation}
These results mean that the magnetic field does not depend on $\zz$, namely it has the form
\begin{equation}
	B^{\xx}=-\tfrac{1}{2}\pd_{\yy}S^{\zz}_{2}(\xx,\yy),\quad B^{\yy}=\tfrac{1}{2}\pd_{\xx}S^{\zz}_{2}(\xx,\yy),\quad B^{\zz}=B^{\zz}_1(\xx,\yy).
\end{equation}
The remaining second order equations from $\{X_1,X_2\}=0$ are equivalent to the conclusion that $s_1^{x_i}$ are independent of $\zz$ as well.

This suggests that $\zz$ might be a cyclic coordinate if it also disappears from the electrostatic potential $W(\xx,\yy,\zz)$. Indeed, the first order equations for $X_2$, 

\begin{align}
\pd_{\xx}m_2=\tfrac{1}{2} S^{\zz}_{2}(\xx,\yy)\pd_{\xx}	S^{\zz}_{2}(\xx,\yy), \quad \pd_{\yy}m_2=\tfrac{1}{2} S^{\zz}_{2}(\xx,\yy)\pd_{\yy}S^{\zz}_{2}(\xx,\yy),\quad \pd_{\zz} m_2=2 \pd_{\zz} W,
\end{align}
imply 
\begin{equation}
	m_2(\xx,\yy,\zz)=\tfrac{1}{4}\left(S^{\zz}_{2}(\xx,\yy)\right)^2+W_3(\zz),\quad W(\xx,\yy,\zz)=W_1(\xx,\yy)+W_3(\zz).
\end{equation}
$W_3(\zz)$ is however immediately eliminated (up to an irrelevant additive constant) using the first order involutivity equations, because coefficients of $p_{\xx}$ and $p_{\yy}$ there read
\begin{equation}
	a\xx \pd_{\zz}W_3(\zz)=0,\quad -a\yy \pd_{\zz}W_3(\zz)=0.
\end{equation}
With $W_3$ eliminated, all equations for $X_2$ are satisfied. We only have to set $m_1=m_1(\xx,\yy)$ to satisfy the involutivity condition $\{X_1,X_2\}=0$ as well.

We therefore have a cyclic coordinate $\zz$, which implies the existence of a first-order integral $\tilde{X}_2=p_{\zz}$ in a suitably chosen gauge, while $X_2=\tilde{X}_2^2$. We stress that this result holds whenever $a\neq0$ regardless of the values of the other parameters $b$ and $c$.

We have yet to solve the equations for $X_1$, starting with \eqref{conds2ord1}. The left-hand sides of the first 3 equations are vanishing. The third one, \eqref{conds2ord13}, gives a simple compatibility 
\begin{equation}\label{3.13}
	\xx \pd_{\yy}S^{\zz}_{2}(\xx,\yy)-\yy\pd_{\xx} S^{\zz}_{2}(\xx,\yy)=0.
\end{equation}
This implies that we may introduce the shifted polar coordinates, which we denote $(r,\phi)$,
\begin{equation}\label{pol s}
	\xx=r\cos(\phi),\quad \yy=r\sin(\phi),
\end{equation}
in which \eqref{3.13} implies $S^{\zz}_{2}(r,\phi)=S^{\zz}_{2}(r).$

The combination $\yy\cdot \eqref{conds2ord15}-\xx\cdot \eqref{conds2ord16}$ determines the dependence of $S_1^{\zz}(r,\phi)$ on $\phi$ explicitly, introducing an arbitrary function $S^{\zz}_{11}(r)$. 
We now eliminate derivatives of $S_1^{\xx}$ and $S^{\yy}_1$ from the equations \eqref{conds2ord11}, \eqref{conds2ord12}, and \eqref{conds2ord14} and subsequently use \eqref{conds2ord15} to eliminate $B^{\zz}_1$, which yields an equation with explicit trigonometric functions of $\phi$ and unknown functions of $r$, $S^{\zz}_{11}(r)$ and $S_2^{\zz}(r)$. Each of the coefficients of $\sin(\phi)$ and $\cos(\phi)$ must vanish separately, we therefore obtain ODEs for the functions of $r$, some of them multiplied by $b$ and/or $c$.

For example the coefficients of $\sin(\phi)$ and $\cos(\phi)^2\sin(\phi)$ are up to nonvanishing constant factors, respectively, 
\begin{equation}\label{3.15}
	b\left(r^{3}\pd_r^4 S^{\zz}_2 \! \left(r \right)+ r^{2}\pd_r^3 S^{\zz}_2 \! \left(r \right)-2 r \pd_r^2 S^{\zz}_2 \! \left(r \right)+2 \pd_r S^{\zz}_2 \! \left(r \right)\right)=0
\end{equation}
and 
\begin{equation}\label{3.16}
	b\left(r^{3}\pd_r^4 S^{\zz}_2 \! \left(r \right)-3 r \pd_r^2 S^{\zz}_2 \! \left(r \right)+3 \pd_r S^{\zz}_2 \! \left(r \right)\right)=0,
\end{equation}
and we get similar equations including the parameter $c$ together with those for $S^{\zz}_{11}(r)$. 

This leads to a split: On the one hand $b\neq0$ or $c\neq0$ for which \eqref{3.15}--\eqref{3.16} and/or similar equations with the parameter $c$ remain ODEs and we always arrive at the solution \eqref{S(r)}, and on the other hand $b=c=0$ for which most of the equations for $S_{11}^{\zz}(r)$ and $S^{\zz}_2(r)$ vanish identically, cf. \eqref{3.28}--\eqref{3.29} below.

\subsection{Parameter $b$ or $c$ nonvanishing}
For concreteness, let us consider $b\neq 0$; the case $b=0,$ $c\neq0$ is very similar and leads to the same final result.

We continue with solving equations \eqref{conds2ord1} as outlined above.

The final form of $S_{11}^{\zz}$ and $S^{\zz}_2$ obtained from  \eqref{3.15}--\eqref{3.16} and other ODEs determining also $S^{\zz}_{11}$ is 
\begin{equation}\label{S(r)}
	S^{\zz}_{11}(r)=\tfrac{1}{2} a s^{\zz}_{11}r^2+s^{\zz}_{13},\quad S^{\zz}_2(r)=s^{\zz}_{24}r^2+s^{\zz}_{21}.
\end{equation}
This result together with the solution of \eqref{conds2ord15} gives us the magnetic field
\begin{equation}
	B^{\xx}(\xx,\yy,\zz)=-s^{\zz}_{24}\yy,\quad B^{\yy}(\xx,\yy,\zz)=s^{\zz}_{24}\xx,\quad B^{\zz}(\xx,\yy,\zz)=\frac{2 \omega s^{\zz}_{24}}{a} \yy+\frac{c \left(10 a^{2}+13 a +4\right) s^{\zz}_{24}}{9 a^{2} \left(a +2\right)}+\frac{s^{\zz}_{11}}{3}.
\end{equation}
Using the remaining equations from \eqref{conds2ord1}, we find also $S^{\xx}_1$ and $S^{\yy}_1$.

We now use the first-order terms in momenta from the involution $\{X_1,X_2\}=0$ again in the polar coordinates \eqref{pol s}. This determines the dependence of $W_1(r,\phi)$ on $\phi$, leaving an undetermined function $W_{1}(r)$.

Next, we consider the zeroth order equation for $X_1$. The relevant term reads
\begin{equation}
	b s^{\zz}_{24} \left(a +\tfrac{1}{2}\right) \psi r^{2}\left( a^4 (a+2)^2 W_{1}'(r)-s^{\zz}_{24} P(r)\right)=0
\end{equation}
with $P(r)$ a polynomial in $r$ stemming from the dependence of $W$ on $\phi$, not explicit for brevity.

Given that $s^{\zz}_{24}=0$ implies constant magnetic field $\vec{B}$ and vanishing electrostatic potential $W=0$, we continue with $s^{\zz}_{24}\neq 0$ and solve for $W_1(r)$. The last undetermined function, $m_1$, is simply integrated from the first order terms in momenta in $\{X_1,H\}=0$.

We hereby summarize our solution to the problem with $a\neq0,b\neq0$ with a suitable redefinition of constants for better readability. The magnetic field $\vec{B}$ depends on 2 arbitrary constants, $b_x=s^{\zz}_{24}$ and $b_{z}=\frac{c \left(10 a^{2}+13 a +4\right) }{9 a^{2} \left(a +2\right)}s^{\zz}_{24}+\frac{1}{3}s^{\zz}_{11}$,

\begin{equation}\label{B sys aneq0}
	B^{\xx}(\xx,\yy,\zz)=-b_x\yy,\quad B^{\yy}(\xx,\yy,\zz)=b_x\xx,\quad B^{\zz}(\xx,\yy,\zz)=\frac{2 \omega b_x}{a} \yy+b_{z},
\end{equation}
and the electrostatic potential $W$ has 3 more constants, $w_1,w_2$ and $w_3$,
\begin{equation}\label{W sys aneq0}
	\begin{split}
		W(\xx,\yy,\zz)={}&\Bigg[	-\frac{\left(a +2\right)^{2} b_x \xx^{4}}{72 a^{2}}+\frac{2 \left(2 a +1\right) b \psi b_x \xx^{3}}{27 a^{3}}+\left(\frac{\left(5 a^{2}-4 a -4\right) b_x \yy^{2}}{12 a^{2}}-\frac{b_{z} \omega \yy}{3 a}\right.\\
		{}&-\left.\frac{b_{z} \left(8 a^{2}+5 a -4\right) c }{9 a^{2} \left(a +2\right)}\right) \xx^{2}+\left(\frac{2 \left(2 a +1\right) b \psi b_x \yy^{2}}{9 a^{3}}-\frac{{2 b b_{z} \omega \psi \yy}}{9 a^{2} \left(a -1\right)}\right) \xx \\
		{}&+\frac{\left(31 a^{2}-20 a -20\right) b_x \yy^{4}}{72 a^{2}}+\left(\frac{2 \left(8 a^{2}+5 a -4\right) c \omega b_x}{27 \left(a +2\right) a^{3}}-\frac{b_{z} \omega}{3 a}\right) \yy^{3}\\{}&+w_3(\xx^2+\yy^2)+w_1 \xx+w_2\yy\Bigg] b_x.
	\end{split}
\end{equation}

Given that the coordinate $\zz\equiv z$ is cyclic (in a suitable gauge), we have a first order integral 
\begin{equation}
	\tilde{X}_2=p_{z}^A+\frac{b_x}{2}(\xx^2+\yy^2)
	\end{equation}
instead of $X_2$ (which is a quadratic polynomial in $\tilde{X}_2$). The integral $X_1$ \eqref{X1 shift} remains quadratic, and we only present the functions in it for better readability:
\begin{align}
	\hspace{-2.45 em}	s_1^{\xx}(\xx,\yy,\zz)
	= {}&
	\left(\frac{\left(-5 a^{2}+4 a +4\right) b_x \yy}{2 a}+b_{z} \omega \right) \xx^{2}-\left(\frac{4 b_x b \psi \left(2 a +1\right) \yy}{3 a^{2}}-\frac{2 \omega b \psi b_{z}}{3 a \left(a -1\right)}\right) \xx\nonumber\\
	\hspace{-2.45 em} {}&-\frac{\left(31 a^{2}-20 a -20\right) b_x \yy^{3}}{6 a}-\left(\frac{2 \left(8 a^{2}+5 a -4\right) c \omega b_x}{3 \left(a +2\right) a^{2}}-3 b_{z} \omega \right) \yy^{2}-6 a w_{3} \yy-3 a w_{2},\nonumber \\
	\hspace{-2.45 em} s_1^{\yy}(\xx,\yy,\zz)
	= {}&
	-\frac{\left(a +2\right)^{2} b_x \xx^{3}}{6 a}+\frac{2 b_x b \psi \left(2 a +1\right) \xx^{2}}{3 a^{2}}+\left(\frac{\left(5 a^{2}-4 a -4\right) b_x \yy^{2}}{2 a}-{2 b_{z} \omega \yy}\right.\label{s1zz}\\
	\hspace{-2.45 em} {}&-\left.\frac{2 \left(8 a^{2}+5 a -4\right) b_{z} c}{3 a \left(a +2\right)}+6 a w_{3}\right) \xx +\frac{2 b_x b \psi \left(2 a +1\right) \yy^{2}}{3 a^{2}}-\frac{2 \omega b \psi b_{z} \yy}{3 a \left(a -1\right)}+3 a w_{1},\nonumber\\ 
	\hspace{-2.45 em} s_1^{\zz}(\xx,\yy,\zz)
	= {}&
	\left(2 \omega b_x \yy -\frac{c \left(a +2\right) b_x}{3 a}+\frac{3 a b_{z}}{2}\right) \xx^{2}+\frac{2 \omega b \psi b_x \yy \xx}{3 a \left(a -1\right)}+2 \omega b_x \yy^{3}\nonumber\\
	\hspace{-2.45 em} {}&-\left(\frac{3 \left(a +1\right) c b_x}{a +2}-\frac{3 a b_{z}}{2}\right) \yy^{2}.\nonumber
\end{align}

The function $m_1$ can be obtained by integrating the first order equations, which in the shifted coordinates \eqref{shift} but without substitution of \eqref{B sys aneq0}--\eqref{s1zz} read
\begin{align}
	\hspace{-2.45 em}		\pd_{\xx}m_1={}&s^{\zz}_1 B^{\yy}-s^{\yy}_1 B^{\zz}+2\left(2\omega\yy-\frac{c(5a^2-a-4)}{3a(a+2)}\right)\pd_{\xx}W
	-2\left(\omega \xx+\frac{b\psi}{3a(a-1)}\right)\pd_{\yy}W+3a\yy \pd_{\zz}W,\notag\\
	\hspace{-2.45 em}	\pd_{\yy}m_1={}&s^{\xx}_1 B^{\zz}-s^{\zz}_1 B^{\xx}-2\omega\left(\xx+\frac{b\psi}{3a(a-1)}\right)\pd_{\xx}W +2c\pd_{\yy}W-3a\xx\pd_{\zz}W,\label{conds1ord1}\\
	\hspace{-2.45 em}	\pd_{\zz}m_1={}&s^{\yy}_1 B^{\xx}-s^{\xx}_1 B^{\yy}	+3a\yy\pd_{\xx}W -3a\xx \pd_{\yy}W-\frac{2c (2a+1)}{a+2}\pd_{\zz}W.\notag
\end{align}
The very long explicit solution for $m_1(\xx,\yy,\zz)$ can be found in Appendix \ref{app}.

There is no additional first order integral commuting with $X_1$ or $X_2$. A more thorough search for additional integrals, i.e. superintegrability, is out of scope of this article.

\subsection{Parameters $b$ and $c$ vanishing}
We note that under these assumptions the shift \eqref{shift} is just the identity transformation, hence we drop the tildes in this subsection.

To recapitulate, we have already determined that the integral $X_2$ from \eqref{X2} reduces to a first order integral $\tilde{X}_2=p_z$ in a suitable gauge. 

We proceed by solving \eqref{conds2ord1}, the second order equations for $X_1$. Proceeding as described under \eqref{pol s}, we obtain conditions on $S^{z}_{11}(r)$ and $S^{z}_2(r)$, which now reduce to only two equations:
\begin{equation}\label{3.28}
	r^{2}\pd_r^3S^{z}_{11}(r)+ r \pd_r^2 S^{z}_{11}(r)-\pd_r S^{z}_{11}(r)=0
\end{equation}
and 
\begin{equation}\label{3.29}
 r^{3} \pd_r^4 S^{z}_2(r)+5 r^{2}\pd_r^3 S^{z}_2(r)+2 r\pd_r^2 S^{z}_2(r) -2\pd_r S^{z}_2(r)=0.
\end{equation}
The solution has more arbitrary constants than before in \eqref{S(r)}, namely
\begin{equation}
	S^{z}_{11}(r)=s^{z}_{11}+s^{z}_{12}\ln(r)+s^{z}_{13} r^2,\quad S^{z}_{2}(r)=s^{z}_{21}r^2+s^{z}_{22}\ln(r)+s^{z}_{23}+\frac{s^{z}_{24}}{r}.
\end{equation}

We solve the equations for $B^{z}_1$ and subsequently for $S^{x}_1$ and $S^{y}_1$ in Cartesian coordinates analogously to the previous case, this time the solution includes $\ln(x^2+y^2)$ and $\arctan(\frac{y}{x}).$ Thus the second order equations are solved and the preliminary form of the magnetic field reads
\begin{equation}
	\begin{split}
	B^{x}(\vec{x}) = {}&
	-3 a \left(a +2\right) \left(s^{z}_{21} +\frac{s^{z}_{22}}{2r^2}- \frac{s^{z}_{24}}{2r^3} \right) y,\\
	B^{y}(\vec{x})= {}&
		3 a \left(a +2\right) \left(s^{z}_{21} +\frac{s^{z}_{22}}{2r^2}- \frac{s^{z}_{24}}{2r^3} \right) x,\\
	B^{z}(\vec{x})= {}&
	\left(a +2\right) \left(\omega\left(6 s^{z}_{21} +\frac{s^{z}_{22}}{r^2}\right) y +3a \left(a +2\right) \left(2 s^{z}_{13} + \frac{s^{z}_{12}}{r^2}\right)\right).
\end{split}
\end{equation}
We now use the first-order terms in momenta from $\{X_1,X_2\}=0$ again in the polar coordinates \eqref{pol s}. This determines the dependence of $W_1(r,\phi)$ on $\phi$, leaving an undetermined function $W_{11}(r)$.

We proceed to the zeroth order equation from $\{X_1,H\}=0$ and separate again into coefficients of $\sin(\phi)^i\cos(\phi)^j$ with different powers $i,j$ ($i=0,1$). One of the coefficients is
\begin{equation}
	\omega^2 \left(2 r s^{z}_{22} -3 s^{z}_{24} \right) \left(2 s^{z}_{21} r^{2}+s^{z}_{22} \right)^{2}=0,
\end{equation}
which (by collecting powers of $r$) implies $s^{z}_{22}=0$.
Two other coefficients yield 
\begin{equation}
s^{z}_{24} (s^{z}_{21})^2=0,\quad \omega (s^{z}_{12})^2 s^{z}_{24}=0,
\end{equation}
implying the split: $s^{z}_{24}=0$ or $s^{z}_{24}\neq 0$ with $s^{z}_{12}=s^{z}_{21}=0$. 

When $s^{z}_{24}=0$, the compatibility condition for the mixed derivatives of $m_1(x,y,z)=M_1(x,y)$ implies that also $s_{12}^{z}=0$ and we find the system \eqref{B sys aneq0} with $b=c=0$. The other case, $s^{z}_{24}\neq0$, gives a cylindrically separable system: Case II in \cite[eq. (7.12)]{Kubu_2021} with constants $A_2=A_3=0$, i.e.
\begin{equation}\label{Case II}
	B^x(\vec{x}) = -A_1\frac{y}{r^3},\quad B^y(\vec{x}) = A_1\frac{x}{r^3} ,\quad B^z(\vec{x}) = 0,\quad W(\vec{x})=\frac{w_1}{r}-\frac{A_1^2}{2r^2}.
\end{equation}
The system \eqref{Case II} admits $p_z$ and $\ell_z$ as integrals and $X_1$ reduces to a parabolic integral $\ell_z p_x$, i.e. it is a superintegrable system separable in cylindrical as well as parabolic cylindrical coordinates.

\section{Parameter $a$ vanishing}\label{sec a0}
We note that $a=0$ implies $\omega=1$. Applying the assumptions on the parameters, the integral $X_1$ of \eqref{X1} reads
\begin{equation}\label{X1a0}
	X_1=-2 \ell_z^A p_x^A-\frac{c}{2}\left(p_x^A\right)^2+ c \left(p_y^A\right)^2-c p_x^A p_z^A+2\beta p_y^A (p_x^A-p_z^A)-\frac{c}{2}\left(p_z^A\right)^2
		+s_1^{x} p_{x}+s_1^{y} p_{y}+s_1^{z} p_{z}+m_1,
\end{equation}
where we have introduced $\beta =b/\sqrt{2}$, which simplifies many formulas below. The other integral $X_2$ retains the form \eqref{X2}.

As was observed in \cite{Marchesiello2022}, the choice $a=0,\omega=1,c=b=0$ leads to the standard parabolic cylindrical case. We are mainly interested in the integrals of the generalized form; we will therefore assume that at least one of the parameters $\beta,c$ is nonvanishing.

The form of $X_2$ implies that we can use the solution to the second order equations \eqref{conds2ord2} and $\nabla\cdot \vec{B}=0$ \eqref{divB} from the $a\neq0$ case, namely \eqref{BS2}, with $\xx=x,\yy=y,\zz=z$.

Looking at three of second order involutivity equations $\{X_1,X_2\}=0$, namely
\begin{equation}\label{a0 comm}
\begin{split}
\hspace{-2 em}	c \pd_zS^x_{21} \! \left(z \right) x -2\left( \beta \pd_zS^x_{21} \! \left(z \right)+ S^x_{21} \! \left(z \right)\right) y -2 \beta \pd_zS^x_{22} \! \left(z \right) +2 S^x_{22} \! \left(z \right) -c \pd_zS^y_2 \! \left(z\right)+3 c S^x_{21} \! \left(z \right)=0,\\
\hspace{-2 em} (\beta \pd_zS^x_{21} \! \left(z \right)+ S^x_{21} \! \left(z \right))x - \beta \pd_zS^y_2 \! \left(z\right) - \beta S^x_{21} \! \left(z \right)=0, \\
\hspace{-2 em} c \pd_zS^x_{21} \! \left(z \right) y -2 \beta S^x_{21} \! \left(z \right)+c \pd_zS^x_{22} \! \left(z \right) +2 S^y_2 \! \left(z \right)=0,
	\end{split}
\end{equation}
it is clear that we have to split, this time into 3 subcases: both $\beta\neq0$ and $c\neq0$, $\beta\neq0$ with $c=0,$ and $c\neq0$ with $\beta=0$. The first and second cases yield a new system, so we will consider them in detail. The last case, $\beta=0,c\neq0$, only contains a reduced version of the system \eqref{a0bcn0B}--\eqref{a0bcn0W} from $\beta \neq0$ with $c\neq0$ together with known superintegrable systems (the constant magnetic field with $W=0$, the Cartesian superintegrable system \cite[Case A.2]{Marchesiello2017} in the form of eq. (50) therein with $\beta=-\alpha$ and $\Omega_2=0$, and the helical undulator from \cite{KMS2022}).

\subsection{Both $\beta$ and $c$ nonvanishing}
This assumption implies that all functions in \eqref{a0 comm} vanish,
\begin{equation}
	S^x_{21}(z)=0,\quad S^x_{22}(z)=0,\quad S^y_2(z)=0
\end{equation}
and the remaining second order equations in $\{X_1,X_2\}=0$ imply that also the functions $s_1^{i}$ do not depend on $z$, $s_1^{x_i}=s_1^{x_i}(x,y)$.

As a consequence of the first order terms in the involutivity equations $\{X_1,X_2\}=0,$ the electrostatic potential separates $W=W_1(x,y)+W_3(z)$, but the zeroth order term in $\{X_1,X_2\}=0$ eliminates $W_3(z)$. Therefore, the coordinate $z$ is cyclic and the integral $X_2$ reduces to a first order one,
\begin{equation}\label{X2 abn0}
	\tilde{X}_2=p_z^A+\tfrac{1}{4}\left(S^z_2(x,y)\right)^2.
\end{equation}

We continue with equations for the integral $X_1$, starting with the quadratic ones, i.e. the analogues of \eqref{conds2ord1} but without the shift \eqref{shift}. The key equation at this moment is
\begin{equation}\label{3.4c}
	2\beta \pd_y S^z_2+c\pd_x S^z_2=0,
\end{equation} 
which implies $S^z_2=S^z_2(y-2\frac{\beta x}{c})$. Next, we integrate equations with $S^x_1$ and $S^z_1$ differentiated by $y$, then find $B^z_{1}$, introducing arbitrary functions $S^x_{11}(x), S^z_{11}(x)$ and $B^z_{11}(y-2\frac{\beta x}{c})$ in the integration.

The remaining leading order equations from $\{H,X_1\}=0$ can be solved for derivatives of $S^x_1(x,y)$. Imposing their compatibility $\pd_{xy}S^x_1=\pd_{yx}S^x_1$ and differentiating with respect to $y$ to eliminate functions of $x$, a coordinate change enables reduction of $S^z_2(y-2\frac{\beta x}{c})$ and $B^z_{11}
(y-2\frac{\beta x}{c})$ to polynomials in their variable. The solution of the remaining second order equations is tedious but straightforward.

The first order involutivity conditions $\{X_1,X_2\}=0$ introduce a function of $(y-2\frac{\beta x}{c})$ into $W$, but the zeroth order equation from $\{X_1,H\}=0$ reduces it to a polynomial or eliminates the magnetic field (we do not consider the latter further).

The remaining calculations are again straightforward and we get (after elimination of constants corresponding to the first order integral $\tilde{X}_2$) a magnetic field $\vec{B}$ depending on 2 constants
$b_y,b_z$
\begin{equation}\label{a0bcn0B}
	B^x(\vec{x})=b_y c,\quad B^y(\vec{x})=2 b_y \beta ,\quad B^z(\vec{x})=6b_y y+b_z,
\end{equation}
and the electrostatic potential $W$ depending on 3 additional parameters $w_1,w_2,w_3,$
\begin{equation}\label{a0bcn0W}
	\begin{split}
		W(\vec{x})={}&\left(-\frac{x^{4}}{2}+2\beta x^{3} -\frac{\left(6 y^{2}+4 \beta^2-c^{2}\right) x^{2}}{2}+6 \left(y +\frac{c}{3}\right) \beta x y -\frac{5 y^{4}}{2}+3 c y^{3} \right) b_y^{2}\\{}&-\left(\left(\frac{\left(2 y +3 c \right) x^{2}}{2}-2 \beta x y +y^{3}\right) {b_z} +\frac{{w_1}}{2}\left(x^{2}+y^{2}\right)+ \beta{w_2}x + {w_3}y \right) {b_y}.
	\end{split}
\end{equation}

The integral $X_2$, retaining the form \eqref{X2}, reduces to the first order one,
\begin{equation}
	\tilde{X}_2=p_z^A+b_y(2\beta x-cy),
\end{equation}
while $X_1$ from \eqref{X1a0} remains quadratic. Its lower order terms are defined by the following functions:
\begin{align}
	s_1^{x}(\vec{x})
	= {}&
	\left(6 b_y y + b_z \right) x^{2}-2\beta \left(6 b_y y + b_y c + b_z \right) x +10 b_y y^{3}-3\left(3 b_y c -b_z \right) y^{2}+ {w_1} y + {w_3},\nonumber \\
	s_1^{y}(\vec{x})
	= {}&
	-2 {b_y} x^{3} +6 {b_y} \beta x^{2} -\left(6 {b_y} y^{2}+2 {b_z} y+\left(4 \beta^2-c^{2}\right) {b_y} +3 c {b_z} + w_1 \right) x\\
	{}&+6 \beta {b_y} y^{2}+2 \beta \left( {b_y} c + {b_z} \right) y - \beta {w_2}
	,\nonumber\\
	s_1^{z}(\vec{x})
	= {}&
	- b_y c x^{2}+2\beta\left(2 b_y y + b_y c + b_z \right) x -3 b_y c y^{2}-\left(4 \beta^2 b_y -3 b_y c^{2}+c b_z \right) y +\tfrac{1}{2}c {w_2}\nonumber,
\end{align}

\begin{align}
	\hspace{-2.45em}	m_1(\vec{x})={}&\left(4 b_y^{2} y +\tfrac{1}{2} c b_y^{2}+b_z b_y \right) x^{4}-4\left[4b_y^{2} \beta y +b_y \beta \left(b_y c +b_z \right) \right] x^{3}\nonumber\\
	{}&+\left[16 b_y^{2} y^{3}-\left(6 c b_y^{2}-8 b_z b_y \right) y^{2}+\left(20 b_y^{2} \beta^2-c^{2} b_y^{2}+4 b_z b_y c +2 b_y w_1 +b_z^{2}\right) y \right.\nonumber\\
	{}&+\left.\frac{\left(12 \beta^2 c -c^{3}\right) b_y^{2}}{2}+\frac{\left(12 \beta^2 b_z +2 b_z c^{2}+c w_1 +2 w_3 \right) b_y }{2}+\frac{3 c b_z^{2}}{2}+\frac{b_z w_1 }{2}\right] x^{2}\nonumber\\
	{}&-\left[32 b_y^{2} \beta y^{3}-2\left( b_y \left(b_y c -8 b_z \right) \beta y+2\beta\left(c^{2}-2 \beta^2\right) b_y^{2}-\left(\left(3 c b_z + w_1 \right) \beta - \beta w_2 \right) b_y - \beta b_z^{2}\right) y \right.\nonumber\\
	{}&-\left.\left(2 b_y c +b_z \right) \beta w_2 +2 b_y w_3 \beta \right] x+12 b_y^{2} y^{5}+\left(7 b_z b_y-\tfrac{37}{2} c b_y^{2} \right) y^{4}\\
	{}&+\left(4 b_y^{2} \beta^2+7 c^{2} b_y^{2}-5 b_z b_y c +2 b_y w_1 +b_z^{2}\right) y^{3}\nonumber\\
	{}&+\left(\frac{\left(8 \beta^2 c -3 c^{3}\right) b_y^{2}}{2}+\frac{\left(4 \beta^2 b_z +b_z c^{2}-2 c w_1 +6 w_3 \right) b_y }{2}+\frac{b_z w_1 }{2}\right) y^{2}\nonumber\\
	{}&-\frac{\left(4 w_2 b_y \beta^2+ \left(4b_y c -2 b_z \right) w_3 + b_y c^{2} w_2 \right) y}{2}. \nonumber
\end{align}

The quadratic terms in  $X_1$ from \eqref{X1a0} can be simplified by a shift of the coordinates to
\begin{equation}
	X_1=\ell_z^A p_x^A-2\beta p_y^A p_z^A - 2c p_x^A p_z^A+\ldots
\end{equation}
Subsequently reducing the system to the $xy$-plane by setting $X_2=\text{const.},$ the last two terms of $X_1$ become first order ones, which leads to a parabolic structure of the integral \cite{Winternitz1967}. The system can be therefore interpreted as an extension of the 2D parabolic case \cite{Pucacco_2005}. On the other hand, the motion in the plane is not completely decoupled unlike e.g. the systems in \cite{Fournier2019}. (See \cite{EscobarRuiz2024} for the related notion of partial integrability applied to various systems including one with a magnetic field.)

\subsection{Parameter $\beta$ nonvanishing, $c$ vanishing}
With this assumption on the parameters, the solution to \eqref{a0 comm} is no longer trivial,
\begin{equation}
	S^x_{21}(z)=s^x_{21}{\mathrm e}^{-\frac{z}{\beta}},\quad S^x_{22}(z)=s^x_{22}{\mathrm e}^{\frac{z}{\beta}},\quad S^y_2(z)=\beta s^x_{21}{\mathrm e}^{-\frac{z}{\beta}},
\end{equation}
and the functions $s_1^{x_i}$ get some terms depending on $z$. 

We proceed with the quadratic equations for $X_1$. Eq. \eqref{3.4c} with $c=0$ implies $S^z_2=S^z_2(x)$, so we avoid the technical difficulties encountered in previous subsection. The other quadratic equations can be solved step by step. To identify splitting of the computation, we need to fix the constants in the solution for $S^z_2(x)$,
\begin{equation}\label{Sz2}
	S^z_2(x)=s^z_{20}+2 s^z_{21}(\beta-x)+\frac{s^z_{22}}{16(\beta-x)^4}+\frac{s^z_{23}}{16(\beta-x)^2}.
\end{equation}

When $s^x_{21}=s^x_{22}=0$, we do not get anything new, only a shifted version of a known superintegrable system \cite[Cases I(b) and I(c)]{Marchesiello2020} corresponding to $s^z_{21}=0$,
\begin{equation}
\begin{split}
		B^x(\vec{x}) ={}& 0, \quad
	B^y(\vec{x})= \frac{b_1}{\left(\beta- x \right)^{3}}, \quad
	B^z(\vec{x}) = 0, \quad
	W(\vec{x})	=
	-\frac{b_1^{2}}{8 \left(\beta- x \right)^{4}}+\frac{w_{1}}{8\left(\beta-x \right)^{2}},
\end{split}
\end{equation}
 and $s^z_{22}=s^z_{23}=0$ leads to the special case of \eqref{a0bcn0B}--\eqref{a0bcn0W} with $c=0$.

When at least one of the constants $s^x_{21}$ and $s^x_{22}$ is nonvanishing, which implies that the integral $X_2$ does not reduce to a first order one, we also have to go through several subcases depending on constants in $S^z_2(z).$ However, only the branch with $s^x_{21}\neq0$ and $s^x_{22}\neq0$ is relevant in the end, all other give reduced version of this branch, i.e. its limits with either $s^x_{21}\to 0$ or $s^x_{22} \to 0$. The final solution for the fields has 4 constants in the magnetic field $\vec{B}$, $b_1=-\frac{s^x_{21}}{2\beta},b_2=-\frac{s^x_{22}}{2\beta},b_y=-\frac{s^z_{21}}{2\beta}$, and $b_z$ which is the constant term in $B^z_1(x,y)$ (the other constants in \eqref{Sz2} vanish),

\begin{equation}\label{B sys a0 exp}
	B^{x}(\vec{x})=b_{1} {\mathrm e}^{-\frac{z}{\beta}} \left(x-\beta \right),\quad B^{y}(\vec{x})=b_{1} {\mathrm e}^{-\frac{z}{\beta}} y-b_{2}{\mathrm e}^{\frac{z}{\beta}}+2\beta b_{y},\quad B^{z}(\vec{x})=2 \beta b_{1} {\mathrm e}^{-\frac{z}{\beta}}+6 b_{y} y+ b_{z}, 
\end{equation}
and the electrostatic potential $W$ has 2 more constants, $w_1,w_3$ (the choice of the notation will be clear soon), 

	\begin{align}
		W(\vec{x})={}&-\tfrac{1}{2}\beta^2 b_{1}^{2}\left( x^{2} -2 \beta x +y^{2}\right){\mathrm e}^{-\frac{2 z}{\beta}}-\beta b_1\left[(x^2-2\beta x+ y^2)(2b_y y +\tfrac{1}{2}b_z)+\tfrac{1}{2}w_3\right]
		{\mathrm e}^{-\frac{z}{\beta}}\nonumber\\
		{}&-\beta b_2\left[(x^2-2\beta x+3 y^2) b_y+b_z y+\tfrac{1}{2}w_1\right]{\mathrm e}^{\frac{z}{\beta}}-\tfrac{1}{2}\beta^2 b_{2}^{2}{\mathrm e}^{\frac{2 z}{\beta}}-\beta^2 b_1 b_2 y \nonumber\\
	{}&+\left[-\tfrac{1}{2}x^{4}+2\beta x^{3} -\left(3 y^{2}+2 \beta^2\right) x^{2}+6 \beta x y^2 -\tfrac{5}{2}y^{4} \right] b_y^{2}\label{W sys a0 exp}\\
	{}&-\left[\left( x^{2} -2 \beta x +y^{2}\right) b_z y +\frac{w_1}{2}\left(x^{2}+y^{2}\right)- \beta{w_1}x + {w_3}y \right] {b_y}.\nonumber
	\end{align}

If we eliminate the exponentials in \eqref{B sys a0 exp} by setting $b_1=b_2=0$, the electrostatic potential $W$ goes to \eqref{a0bcn0W} where we set $c=0$, $w_2=-w_1$ and $b_1=b_y$.

This time, both integrals $X_1$ of the form \eqref{X1a0} and $X_2$ as in \eqref{X2} are quadratic, the functions determining their lower order terms read
\begin{align}
	s_1^x(\vec{x})= {}&
	4 \beta b_{2}y {\mathrm e}^{\frac{z}{\beta}}+2 \beta b_{1} \left(x^{2}-2 \beta x+2 y^{2}\right) {\mathrm e}^{-\frac{z}{\beta}}+2\left(3 x^{2} -6\beta x+5 y^{2}\right) b_{y}y \nonumber\\
	{}&+\left(x^{2}-2 \beta x+3 y^{2}\right) b_{z}+w_1 y+w_3,\\
	 s_1^y(\vec{x})= {}&-
	\left[2 \beta b_{2}{\mathrm e}^{\frac{z}{\beta}}+2 \beta b_{1} y {\mathrm e}^{-\frac{z}{\beta}}+ 2 b_{y}(x^2-2\beta x+3y^2)+2b_z y+w_1\right] \left(x-\beta\right),\nonumber\\
	s^z_1(\vec{x})={}&
	2\beta\left[b_{1} \beta {\mathrm e}^{-\frac{z}{\beta}}+ 2b_{y}y+ b_{z} \right] \left(x-\beta \right)\nonumber,
\end{align}

	\begin{align}
		\hspace{-2.45em}	m_1(\vec{x})={}&2\beta^2 b_{2}^{2} y {\mathrm e}^{\frac{2z}{\beta}}+b_2\beta \left[2\left(3x^2-6\beta x+5y^2+2\beta^2\right) b_y y+b_z\left(x^2-2\beta x+3y^2 +2\beta^2\right) +w_1 y+w_3\right] {\mathrm e}^{\frac{z}{\beta}}\nonumber\\
		{}&+\beta b_1 \left[2 b_y(x^2-2\beta x+y^2)(x^2-2\beta x+5y^2+\beta^2)+(x^2-2\beta x +y^2)3 b_z y+\right.\nonumber\\
		{}&+\left. w_1\left((x-\beta)^2+y^2\right)+w_{3} y\right]
		{\mathrm e}^{-\frac{z}{\beta}}+2 y \beta^2 b_{1}^{2} \left(x^{2}+y^{2}-2 \beta x\right) {\mathrm e}^{-\frac{2z}{\beta}}\\
		{}&+ \left(x^{2} -2 \beta x+y^{2}\right)\left[4b_y^2 y \left((x-\beta)^2+3y^2\right) +\left(x^2-2\beta x+7y^2+2\beta^2\right) b_y b_z\right.\nonumber\\
		{}&+\left.b_z^{2}y+\tfrac{1}{2}b_z w_1\right]+2 \beta^{2} b_1 b_2 \left(x^{2} -2 \beta x+2 y^{2}\right)\nonumber\\
			{}&+2 b_y w_1 \left((x-\beta)^2+y^2\right) y+[(x^2-2\beta x+3 y^2)b_y+b_z y] w_3, \nonumber
	\end{align}

\begin{align}
s_2^x(\vec{x}) ={}&	
-2 \beta b_{1} {\mathrm e}^{-\frac{z}{\beta}} y -2 \beta b_{2} {\mathrm e}^{\frac{z}{\beta}},\quad
s_2^y(\vec{x})= 
2 \beta b_{1} {\mathrm e}^{-\frac{z}{\beta}} \left(x-\beta \right), \quad
s_2^z(\vec{x})= 4\beta b_{y} \left(x-\beta \right),
\end{align}

\begin{equation}
	\begin{split}
		m_2(\vec{x})={}&-2 \beta^2 b_{2}^{2} {\mathrm e}^{\frac{2z}{\beta}}-2\beta b_2 \left[\left(x^2-2\beta x+3y^2-2\beta^2 \right) b_y+b_z y+\tfrac{1}{2} w_{1}\right]{\mathrm e}^{\frac{z}{\beta}}\\
		{}&-4\beta b_1\left[\left((x-\beta)^2+y^2\right)b_y y +\tfrac{1}{4}\left(x^2-2\beta x+y^2\right)b_z+\tfrac{1}{4} w_{3}\right] {\mathrm e}^{-\frac{z}{\beta}}\\
		{}&-\beta^2 b_{1}^{2} \left(\beta^2-4 \beta x +2 x^{2}+2 y^{2}\right){\mathrm e}^{-\frac{2z}{\beta}}+{4\beta^2 b_{y}^{2} \left(x^{2}-2 \beta x \right)} -4 \beta^2 b_{1} b_{2} y.
		\end{split}
	\end{equation}

	\section{Conclusions}\label{sec cocl}
	In this article on quadratically integrable systems with magnetic field possessing integrals of motion not connected to separability, we have classified such systems in the generalized parabolic cylindrical class \cite[class (j)]{Marchesiello2022}.
	
	Depending on the values of the parameters $a,b,c$, there are 3 families of such systems not known before: When $a\neq 0$ and the integral of motion $X_1$ \eqref{X1 shift} has 2 terms depending on the angular momentum, we have found the system \eqref{B sys aneq0}--\eqref{W sys aneq0} with linear magnetic field $\vec{B}$ and quartic scalar potential $W$. When $a=0$, we have 2 families: The first one allowing both remaining parameters $b$ and $c$ nonvanishing, \eqref{a0bcn0B}--\eqref{a0bcn0W}, also has a rather simple linear $\vec{B}$ and quartic potential, while the second one with $b\neq 0$ and $c=0$, \eqref{B sys a0 exp}--\eqref{W sys a0 exp}, has the fields depending on exponential functions of the coordinate $z$.
	
	Let us consider the limit of vanishing magnetic field. All three systems reduce to the free motion, thus they are obviously separable in every choice of orthogonal coordinates in the Euclidean space $\mathbb{E}_3$. This behavior is consistent with previous results on nonstandard integrable systems, which in the limit $\vec{B}\to 0$ always separate in Cartesian or cylindrical coordinates, i.e. in subgroup-type coordinate systems, whether they extend this type of coordinates or not \cite{KMS2022,Hoque2023,Marchesiello2022}. On the other hand, the limit for standard systems typically does not eliminate the potential $W$ and leads to separation only in the corresponding coordinates, see e.g. the spherical case \cite{Marchesiello2018Sph} (which does not admit extensions \cite{KMS2022}). We therefore confirm the not well understood disparity between subgroup- and non-subgroup-type coordinates, which shall be investigated further.
	 
	Given the computational complexity in classifying the parabolic case in 2D \cite{Pucacco_2005}, the 3D cylindrical parabolic systems, here corresponding to the limit $a,b,c\to 0$, have not yet been classified. We therefore investigate this limit of the systems found above expecting to find yet unknown integrable systems of that type.
	
	In the case $a\neq 0$, \eqref{B sys aneq0}--\eqref{W sys aneq0}, we have to take the limit $b,c\to0$ first, redefine $b_x=a \tilde{b}_x, w_i=\frac{\tilde{w}_i}{a}$ and only then $a\to0$. This leads to the following fields 
	\begin{equation}\label{limit param}
	\begin{split}
			B^{x}(\vec{x})={}&0,\quad B^{y}(\vec{x})=0,\quad B^{z}(\vec{x})=2 \tilde{b}_x y+b_{z},\\
	W(\vec{x})={}&\left[-\tfrac{1}{18}\tilde{b}_x\left(x^{4} +6 x^{2} y^{2}+5 y^{4} \right) -\tfrac{1}{3}b_z\left(x^{2} +y^{2}\right) y +\tilde{w}_3\left(x^{2}+y^{2}\right)+ \tilde{w}_1x + \tilde{w}_2y\right] \tilde{b}_x,
		\end{split}
	\end{equation}
	yielding a parabolic cylindrical system with magnetic fields that can be seen as a trivial extension of a parabolic 2D system. (In this limit, the shift \eqref{shift} is vacuous.) As far as we know this is a new system.
	
	The limit $b,c\to 0$ (thus $\beta\to0$) for the system \eqref{a0bcn0B}--\eqref{a0bcn0W} does not entail such technical difficulties. We get the system \eqref{limit param} but with $\tilde{w}_1=0,\tilde{w}_2=\tfrac{1}{3}w_3,\tilde{w}_3=\tfrac{1}{6}w_1$ and $\tilde{b}_x\to 3 b_y$. 
	On the other hand, the limit for system \eqref{B sys a0 exp}--\eqref{W sys a0 exp} with $b_1\neq0$ or $b_2\neq0$ diverges due to the exponentials. When the exponentials of \eqref{B sys a0 exp}--\eqref{W sys a0 exp} vanish, i.e. $b_1=b_2=0$, the system differs from \eqref{a0bcn0B}--\eqref{a0bcn0W} only by the absence of the terms involving $c$, hence the limit $b,c\to0$ again yields the system \eqref{limit param} with the redefinitions above.
	
	\section*{Acknowledgment}
	Computations in this paper were performed using computer algebra software Maple\textsuperscript{TM} by Maplesoft, a division of Waterloo Maple, Inc., Waterloo, Ontario, and Mathematica\textsuperscript{TM} by Wolfram Research, Inc., Champaign, IL.
	
	OK was supported by the Grant Agency of the Czech Technical University in Prague, grant No. SGS22/178/OHK4/3T/14.
	
	\appendix
	\section{Appendix}\label{app}
	In this brief appendix we present the explicit solution of equations \eqref{conds1ord1}.

	\begin{align}
		\hspace{-2.45em} m_1(\vec{x})={}&\left(\frac{b_x^{2} \left(a +2\right) \omega \yy}{9 a^{2}}-\frac{c \left(4 a^{2}+a +4\right) b_x^{2}}{108 a^{3}}+\frac{b_{z} \left(a +2\right) b_x}{12 a}\right)\left(a +2\right) \xx^{4}\nonumber\\
		{}&+\left(-\frac{\left(23 a^{2}-16 a -16\right) b \psi \omega b_x^{2} \yy}{27 a^{3} \left(a -1\right)}+\frac{4 \psi \omega^2 b c \left(5 a +4\right) b_x^{2}}{81 a^{4} \left(a +2\right)}-\frac{8 b \psi \left(a +\frac{1}{2}\right) b_{z} b_x}{9 a^{2}}\right) \xx^{3}\nonumber\\
		{}&+\left[-\frac{2 \left(7 a^{2}-8 a -8\right) \omega b_x^{2} \yy^{3}}{9 a^{2}}+\left(\frac{c \left(2 a +1\right) \left(a -2\right) b_x^{2}}{3 a^{2}}-\frac{b_{z} \left(23 a^{2}-16 a -16\right) b_x}{6 a}\right) \yy^{2}\right.\nonumber\nonumber\\
		{}&+\left.\left(-4 \omega b_x w_{3}+\frac{8 \left(a +2\right) \left(a +\frac{1}{2}\right) b^{2} \omega b_x^{2}}{27 a^{4}}+\frac{2 c \left(13 a^{2}+4 a -8\right)  \omega b_x b_{z}}{9 a^{2} \left(a +2\right)}+b_{z}^{2} \omega \right) \yy \right.\nonumber\nonumber\\
		{}&+\left.\left(-\frac{2 \left(5 a^{2}-a -4\right) c b_x}{3 a \left(a +2\right)}-3 a b_{z}\right) w_{3}-b_x \omega w_{2}\right.\nonumber\nonumber\\
		{}&+\left.\frac{2 \left(2 a^{4} b^{2}+40 a^{4} c^{2}+13 a^{3} b^{2}+17 a^{3} c^{2}+30 a^{2} b^{2}-57 a^{2} c^{2}+28 a b^{2}-16 a c^{2}+8 b^{2}+16 c^{2}\right) b_x b_z}{27 a^{3} \left(a +2\right)^{2}}\right.\nonumber\\
		{}&+\left.\frac{b_{z}^{2} \left(8 a^{2}+5 a -4\right) c}{3 a \left(a +2\right)}\right] \xx^{2}\nonumber\\
		{}&+\left[{-\frac{\psi \left(55 a^{2}-32 a -32\right) b \omega b_x^{2} \yy^{3}}{27 a^{3} \left(a -1\right)}}-\frac{4 \psi \left(2 a +1\right) b b_x \left(4 a b_{z}-c b_x\right) \yy^{2}}{9 a^{3}}\right.\nonumber\\
		{}&+\left.\left(-\frac{4 b \psi \omega b_x w_{3}}{3 a \left(a -1\right)}-2 b_x \omega w_{1}+\frac{20 \left(a +\frac{4}{5}\right) b \psi c \omega  b_x b_{z}}{27 a^{3} \left(a +2\right)}+\frac{2 b \psi b_{z}^{2} \omega}{3 a \left(a -1\right)}\right) \yy\right.\nonumber\\
		{}&-\left.\frac{2 b b_x \psi \omega w_{2}}{3 a \left(a -1\right)}+\left(\frac{2 c b_x \left(-5 a^{2}+a +4\right)}{3 \left(a +2\right) a}-3 a b_{z}\right) w_{1}\right]\xx	 -\frac{\left(5 a^{2}-4 a -4\right) \omega b_x^{2} \yy^{5}}{3 a^{2}}\nonumber\\
		{}&+\left(\frac{c \left(196 a^{4}+39 a^{3}-312 a^{2}-52 a +48\right) b_x^{2}}{36 a^{3} \left(a +2\right)}-\frac{b_{z} \left(47 a^{2}-28 a -28\right) b_x}{12 a}\right) \yy^{4}\nonumber\nonumber\\
		{}&+\left(-4 b_x \omega w_{3}+\frac{4 \left(2 a^{3} b^{2}+24 a^{3} c^{2}+9 a^{2} b^{2}+15 a^{2} c^{2}+12 a b^{2}-12 a c^{2}+4 b^{2}\right) \omega b_x^{2}}{81 \left(a +2\right) a^{4}}\right.\nonumber\\
		{}&-\left.\frac{22 \left(a^{2}+a -\frac{4}{11}\right) c \omega b_x b_{z}}{9 a^{2} \left(a +2\right)}+b_{z}^{2} \omega \right) \yy^{3}\nonumber\\
		{}&+\left((2 c b_x-3 a b_{z}) w_{3}-3 b_x \omega w_{2}+\frac{4 \left(a +2\right) \left(a +\frac{1}{2}\right) b^{2} b_x b_z}{27 a^{3}}\right) \yy^{2}\nonumber\\
		{}&-\left( \left(3a b_{z}-2 c b_x\right) w_{2}+\frac{2 b b_x \psi \omega w_{1}}{3 a \left(a -1\right)}\right) \yy	\nonumber,
	\end{align}
	where we recall that $\omega = \sqrt{1+a-2 a^2}$ and $\psi=\sqrt{-a^2 - a + 2}$.

\bibliographystyle{plain}
\bibliography{ParCyl}

\begin{thebibliography}{10}

\bibitem{BALAL2020163895}
N.~Balal, V.~L. Bratman, and E.~Magory.
\newblock New varieties of helical undulators.
\newblock {\em Nucl. Instrum. Methods Phys. Res.}, 971:163895, 2020.

\bibitem{Benenti2001}
S.~Benenti, C.~Chanu, and G.~Rastelli.
\newblock Variable separation for natural {H}amiltonians with scalar and vector
  potentials on {R}iemannian manifolds.
\newblock {\em J. Phys. A}, 42(5):2065--2091, 2001.

\bibitem{Berube2004}
J.~B{\'e}rub{\'e} and P.~Winternitz.
\newblock Integrable and superintegrable quantum systems in a magnetic field.
\newblock {\em J. Math. Phys.}, 45(5):1959--1973, 2004.

\bibitem{CampoamorStursberg2013}
R.~Campoamor-Stursberg, J.~F. Cariñena, and M.~F. Rañada.
\newblock Higher-order superintegrability of a {H}olt related potential.
\newblock {\em J. Phys. A: Math. Theor.}, 46(43):435202, 2013.

\bibitem{EscobarRuiz2024}
A.~M. Escobar-Ruiz and R.~Azuaje.
\newblock On particular integrability in classical mechanics.
\newblock {\em J. Phys. A: Math. Theor.}, 57(10):105202, 2024.

\bibitem{EscobarRuiz2017}
A.~M. Escobar-Ruiz, J.~C. L{\'{o}}pez~Vieyra, and P.~Winternitz.
\newblock Fourth order superintegrable systems separating in polar coordinates.
  {I}. {E}xotic potentials.
\newblock {\em J. Phys. A: Math. Theor.}, 50(49):495206, 2017.

\bibitem{EscobarRuiz2018}
A.~M. Escobar-Ruiz, P.~Winternitz, and {\.{I}}.~Yurdu{\c{s}}en.
\newblock General {N}th-order superintegrable systems separating in polar
  coordinates.
\newblock {\em J. Phys. A: Math. Theor.}, 51(40):40LT01, 2018.

\bibitem{Fournier2019}
F.~Fournier, L.~\v{S}nobl, and P.~Winternitz.
\newblock Cylindrical type integrable classical systems in a magnetic field.
\newblock {\em J. Phys. A}, 53(8):085203, 2020.

\bibitem{Fris1965}
J.~Fri\v{s}, V.~Mandrosov, {Ya. A.} Smorodinsky, M.~Uhl\'{\i}\v{r}, and
  P.~Winternitz.
\newblock On higher symmetries in quantum mechanics.
\newblock {\em Phys. Lett.}, 16(3):354--356, 1965.

\bibitem{HeinIld}
T.~Heinzl and A.~Ilderton.
\newblock Superintegrable relativistic systems in spacetime-dependent
  background fields.
\newblock {\em J. Phys. A}, 50(34):345204, 14, 2017.

\bibitem{Hoque2023a}
F.~Hoque and L.~\v{S}nobl.
\newblock Family of nonstandard integrable and superintegrable classical
  {H}amiltonian systems in non-vanishing magnetic fields.
\newblock {\em J. Phys. A}, 56:165203, 2023.

\bibitem{Hoque2023}
M.~F. Hoque, O.~Kub{\r{u}}, A.~Marchesiello, and L.~{\v{S}}nobl.
\newblock New classes of quadratically integrable systems with velocity
  dependent potentials: non-subgroup type cases.
\newblock {\em Eur. Phys. J. Plus}, 138(9):845, 2023.

\bibitem{Hoque2018a}
M.~F. Hoque, I.~Marquette, and Y.-Z. Zhang.
\newblock On superintegrable monopole systems.
\newblock {\em J. Phys. Conf. Ser.}, 965:012018, 2018.

\bibitem{KMW76}
E.~G. Kalnins, W.~Miller, Jr., and P.~Winternitz.
\newblock The group {${\rm O}(4)$}, separation of variables and the hydrogen
  atom.
\newblock {\em SIAM J. Appl. Math.}, 30:630--664, 1976.

\bibitem{Kress2023}
J.~Kress, K.~Sch\"{o}bel, and A.~Vollmer.
\newblock An algebraic geometric foundation for a classification of
  second-order superintegrable systems in arbitrary dimension.
\newblock {\em J. Geom. Anal.}, 33(11):360, 2023.

\bibitem{KMS2022}
O.~Kub\r{u}, A.~Marchesiello, and L.~\v{S}nobl.
\newblock New classes of quadratically integrable systems in magnetic fields:
  The generalized cylindrical and spherical cases.
\newblock {\em Ann. Phys.}, 451:169264, 2023.

\bibitem{Kubu_2021}
O.~Kub{\r{u}}, A.~Marchesiello, and L.~{\v{S}}nobl.
\newblock Superintegrability of separable systems with magnetic field: the
  cylindrical case.
\newblock {\em J. Phys. A: Math. Theor.}, 54(42):425204, 2021.

\bibitem{Makarov1967}
{A. A.} Makarov, {Ya. A.} Smorodinsky, Kh. Valiev, and P.~Winternitz.
\newblock A systematic search for nonrelativistic systems with dynamical
  symmetries.
\newblock {\em Nuovo Cimento A Series}, 10:1061--1084, 1967.

\bibitem{Marchesiello2015a}
A.~Marchesiello, S.~Post, and L.~{\v{S}}nobl.
\newblock Third-order superintegrable systems with potentials satisfying only
  nonlinear equations.
\newblock {\em J. Math. Phys.}, 56(10):102104, 2015.

\bibitem{Marchesiello2022}
A.~Marchesiello and L.~{\v{S}}nobl.
\newblock Pairs of commuting quadratic elements in the universal enveloping
  algebra of {E}uclidean algebra and integrals of motion.
\newblock {\em J. Phys. A}, 55(14):145203, 2022.

\bibitem{Marchesiello2017}
A.~Marchesiello and L.~\v{S}nobl.
\newblock Superintegrable 3{D} systems in a magnetic field corresponding to
  {C}artesian separation of variables.
\newblock {\em J. Phys. A}, 50(24):245202, 24, 2017.

\bibitem{Marchesiello2020}
A.~Marchesiello and L.~\v{S}nobl.
\newblock Classical superintegrable systems in a magnetic field that separate
  in {C}artesian coordinates.
\newblock {\em SIGMA Symmetry Integrability Geom. Methods Appl.}, 16, 2020.
\newblock 015, 35 pages.

\bibitem{Marchesiello2018Sph}
A.~Marchesiello, L.~\v{S}nobl, and P.~Winternitz.
\newblock Spherical type integrable classical systems in a magnetic field.
\newblock {\em J. Phys. A}, 51(13):135205, 24, 2018.

\bibitem{McSween2000}
E.~McSween and P.~Winternitz.
\newblock Integrable and superintegrable {H}amiltonian systems in magnetic
  fields.
\newblock {\em J. Math. Phys.}, 41(5):2957--2967, 2000.

\bibitem{Miller2018}
W.~Miller, Jr., E.~G. Kalnins, and J.~M. Kress.
\newblock {\em Separation of Variables and Superintegrability: The Symmetry of
  Solvable Systems}.
\newblock Bristol: {IOP} Publ., 2018.

\bibitem{MPW81}
W.~Miller, Jr., J.~Patera, and P.~Winternitz.
\newblock Subgroups of {L}ie groups and separation of variables.
\newblock {\em J. Math. Phys.}, 22(2):251--260, 1981.

\bibitem{Miller2013}
W.~Miller, Jr., S.~Post, and P.~Winternitz.
\newblock Classical and quantum superintegrability with applications.
\newblock {\em J. Phys. A}, 46(42):423001, 97, 2013.

\bibitem{Post2011}
S.~Post and P.~Winternitz.
\newblock A nonseparable quantum superintegrable system in 2{D} real
  {E}uclidean space.
\newblock {\em J. Phys. A: Math. Theor.}, 44(16):162001, 2011.

\bibitem{Pucacco_2005}
G.~Pucacco and K.~Rosquist.
\newblock Integrable {H}amiltonian systems with vector potentials.
\newblock {\em J. Math. Phys.}, 46(1):012701, 2005.

\bibitem{Reyes2022}
D.~Reyes, P.~Tempesta, and G.~Tondo.
\newblock Classical multiseparable {H}amiltonian systems, superintegrability
  and {H}aantjes geometry.
\newblock {\em Commun. Nonlinear Sci.}, 104:106021, 2022.

\bibitem{Shapovalov1972}
V.~N. Shapovalov, V.~G. Bagrov, and A.~G. Meshkov.
\newblock Separation of variables in the stationary {S}chr\"{o}dinger equation.
\newblock {\em Sov. Phys. J.}, 15(8):1115--1119, 1972.

\bibitem{Tempesta2021}
P.~Tempesta and G.~Tondo.
\newblock Haantjes algebras of classical integrable systems.
\newblock {\em Ann. Mat. Pur. Appl.}, 201(1):57--90, 2021.

\bibitem{Winternitz1967}
P.~Winternitz, Ya.~A. Smorodinsky, M.~Uhl\'{\i}\v{r}, and I.~Fri\v{s}.
\newblock Symmetry groups in classical and quantum mechanics.
\newblock {\em Sov. J. Nucl. Phys.}, 4:444--450, 1967.

\bibitem{Zhang}
P.-M. Zhang, L.-P. Zou, {P. A.} Horvathy, and {G. W.} Gibbons.
\newblock Separability and dynamical symmetry of quantum dots.
\newblock {\em Ann. Phys.}, 341:94--116, 2014.

\end{thebibliography}
\end{document}